\documentclass[aps,prb,showpacs,twocolumn,superscriptaddress]{revtex4}

\usepackage{graphicx}
\usepackage{amssymb}
\usepackage[ansinew]{inputenc}

\begin{document}

\title{Scanning tunneling microscopy and kinetic Monte Carlo investigation \\ of Cesium superlattices on Ag(111)}
\author{M. Ziegler}
\author{J. Kr\"{o}ger}\email{kroeger@physik.uni-kiel.de}
\author{R. Berndt}
\affiliation{Institut f\"{u}r Experimentelle und Angewandte Physik, Christian-Albrechts-Universit\"{a}t zu Kiel, D-24098 Kiel, Germany}
\author{A. Filinov}
\affiliation{Institut f\"{u}r Theoretische Physik und Astrophysik, Christian-Albrechts-Universit\"{a}t zu Kiel, D-24098 Kiel, Germany}
\affiliation{Institute of Spectroscopy RAS, Moscow region, Troitsk, 142190, Russia}
\author{M. Bonitz}
\affiliation{Institut f\"{u}r Theoretische Physik und Astrophysik, Christian-Albrechts-Universit\"{a}t zu Kiel, D-24098 Kiel, Germany}

\begin{abstract}
Cesium adsorption structures on Ag(111) were characterized in a low-temperature
scanning tunneling microscopy experiment. At low coverages, atomic resolution
of individual Cs atoms is occasionally suppressed in regions of
an otherwise hexagonally ordered adsorbate film on terraces. Close to step
edges Cs atoms appear as elongated protrusions along the step edge direction.
At higher coverages, Cs superstructures with atomically resolved hexagonal
lattices are observed. Kinetic Monte Carlo simulations model the observed
adsorbate structures on a qualitative level.
\end{abstract}

\pacs{68.35.-p,68.37.-d,68.37.Ef,68.43.Fg}

\maketitle

\section{Introduction}
Alkali metal adsorption on metal surfaces has been studied intensely since
many years (for a review the reader is referred to Refs.\,\onlinecite{rdd_96,hpb_89}).
With their single $s$ electron at the outermost atomic shell these metals
are considered as simple and much research is devoted to the analysis of the
interplay between their mutual interaction and the interaction with the
hosting substrate. \cite{rdd_96}
Previous structural studies of Ag(111)-Cs focussed on adsorbate superstructures
at higher coverages. For instance, a $p(2\times 2)$ structure along with
evidence of some disordered and possibly incommensurate phases was found.
\cite{rgr_85,cca_85} Bond length changes between the Cs adsorption layer
(adlayer) and the Ag(111) substrate surface were studied by surface-extended
X-ray absorption fine structure also for higher coverages. \cite{gla_88}
Leatherman and Diehl provided a structural analysis of Ag(111)-Cs
for various coverages and temperatures using low-energy electron diffraction.
\cite{gle_96} In particular, for very low coverages they found ring-like
diffraction patterns which were assigned to disordered or fluid overlayer
phases. This fluid phase appears to be a common arrangement of alkali metal
films on metal surfaces at low coverage. \cite{sal_83,wcf_88,zyl_91,tvh_06}
It is usually argued that at low coverages the dipole-dipole repulsion between
the adsorbed alkali metal atoms dominates the alkali-substrate interaction
resulting in a structure with no long-range order, yet with a typical distance
between the adsorbed atoms (adatoms). Besides these structural properties,
alkali metal-covered surfaces exhibit intriguing electronic properties resulting
from quantum size effects: Quantum well states confined to ultrathin films
of alkali metals are investigated experimentally as well as theoretically.
\cite{sal_88,nfi_93,tfa_95,aca_97,mba_97,sog_99,bhe_00,tch_00,mni_01,jkl_02,aka_05,cco_05,jkr_05}

While the dipole-dipole interaction presents a direct coupling between
Cs adatoms, the substrate may influence the Cs-Cs interaction in an indirect
way. Indirect interaction between two atoms was first investigated theoretically
by Kouteck\'{y} \cite{jko_58} and then by Grimley, \cite{tgr_67} Newns, \cite{dne_69}
and Einstein and Schrieffer. \cite{tei_73} In an early field ion microscopy
experiment evidence for indirect interaction between Re atoms adsorbed on
W(110) was reported by Tsong. \cite{tts_73} In particular, it was found that
the interaction energy exhibits an oscillatory behavior as a function of the
separation distance between the adatoms. Lau and Kohn \cite{kla_78} then predicted
a long-range and oscillatory interaction between atoms mediated by Friedel
oscillations of a two-dimensional electron gas. The first scanning tunneling
microscopy (STM) experiment evidencing this type of interaction was reported
by Brune {\it et al.}\,\cite{hbr_90} for carbon atoms adsorbed on Al(111).
Similar experiments were then performed for benzene molecules on Cu(111),
\cite{mka_96} sulfur atoms on Cu(111), \cite{ewa_98} copper atoms on Cu(111),
\cite{jre_00} cobalt atoms on Cu(111) and Ag(111), \cite{nkn_02} and for Ce
atoms adsorbed on Ag(111). \cite{fsi_04} These experiments reveal that the
adatom-adatom distance is predominantly influenced by the surface state-mediated
interaction. In particular, the mutual distance is reported to be
$\lambda_{\text{F}}/2$ where $\lambda_{\text{F}}$ denotes the Fermi wavelength
of the involved surface state. This interaction may lead to superlattices on
surfaces \cite{fsi_04} or to confinement-induced adatom self-organization in
quantum corrals. \cite{neg_06,vss_06}

Here, we report a low-temperature STM experiment on Cs adsorbed on Ag(111)
at various coverages. At low coverages, substrate terraces are covered by Cs
adatoms which locally exhibit hexagonally ordered domains. Long-range hexagonal
order of the adlayer, however, does not exist owing to regions where
individual Cs adatoms are not resolved in STM images. Similarly, at and close
to step edges individual Cs atoms are not resolved and appear as almost
continuous rows running parallel to the step edges. At higher coverages
the characteristic smearing of adsorbate structures on terraces and at step
edges disappears and individually resolved adatoms form superlattices with
long-range hexagonal order. Kinetic Monte Carlo simulations model the
experimental situation on a qualitative level. We found that both the surface
state-mediated interaction between Cs adatoms and the dipole-dipole interaction
are equally important, in particular for low coverages, and have to be included
in the model. The simulations reveal that loss of atomic resolution is likely
due to the low adsorbate coverage matching an incommensurate superstructure.
At higher coverages, as a result of a reduced adatom-adatom distance, the
mutual dipole repulsion is strong enough to stabilize a superlattice with
long-range hexagonal symmetry and to suppress adatom diffusion.

\section{Experiment}
Experiments were performed using a custom-built scanning tunneling microscope
operated at $7\,\text{K}$ and in ultrahigh vacuum with a base pressure of
$10^{-9}\,\text{Pa}$. The Ag(111) surface as well as chemically etched tungsten
tips were cleaned by argon ion bombardment and annealing. Cesium was deposited
at room temperature from commercial dispensers \cite{sae_gt} at a rate of
$\approx 0.05\,\text{ML}\,\text{min}^{-1}$ as monitored by a quartz microbalance
and judged from the deposition time and corresponding STM images. We define
a monolayer (ML) as one Cs atom per Ag atom. All STM images presented in this
work were obtained in the constant-current mode with voltages applied to the
sample.

\section{Simulations}
We model aspects of our results by using the kinetic Monte Carlo method \cite{dco_65,wyo_66}
which got a firm theoretical background by Fichthorn and Weinberg. \cite{fic_91}
Recently, the kinetic Monte Carlo method has been used to study the
self-organization of adatoms due to surface state-mediated interactions.
\cite{neg_06,vss_06} The substrate lattice is modeled by
finite hop rates of adatoms between adjacent lattice sites $r_i$ and $r_j$.
The hop rate $\nu_{ij}$ is described by an Arrhenius law according to
\begin{equation}
  \nu_{ij}=\nu_{0}\,\exp\left(-E_{ij}/\text{k}_{\text{B}}T\right),
  \label{hop}
\end{equation}
where $T$ is the substrate temperature, $\text{k}_{\text{B}}$ Boltzmann's
constant and $\nu_{0}$ the attempt frequency. For the hopping barrier between
lattice sites $i$ and $j$ we use $E_{ij}=E_{\text{D}}+0.5\,(E_j-E_i)$ where
$E_{\text{D}}$ is the diffusion barrier height for an isolated adatom and the
$E_{j(i)}$ describe the interaction of atom $j$ ($i$) with all the other
atoms. \cite{fic_03} The total interaction is the sum of the dipole-dipole
coupling
\begin{equation}
  E_i^{\text{d}} = \frac{1}{4\pi\epsilon_0}\sum_{j\neq i}\frac{p^2}{|r_j - r_i|^3},
  \label{dip_int}
\end{equation}
where $p$ is the dipole moment of an individual Cs adatom, $\epsilon_0$ the
dielectric constant, and the surface state-mediated interaction \cite{phy_00}
\begin{equation}
  E_i^{\text{s}} = A_0\,E_0\left[\frac{2\sin(\delta_0)}{\pi}\right]^{2}\,\sum_{j\neq i}\frac{\sin(2k_{\text{F}}|r_j - r_i|+2\delta_0)}{(k_{\text{F}}|r_j - r_i|)^2}
  \label{SS_Pot}
\end{equation}
with $A_0$ denoting the scattering amplitude, $E_0$ the surface state binding
energy, $\delta_0$ the scattering phase shift, and $k_{\text{F}}$ the Fermi
wave vector.

Our idea is that regions of the adsorbate lattice that exhibit loss of atomic
resolution are due to an enhanced mobility of the adatom. It is thus desirable
to access the probability of finding a particle at a given lateral coordinate.
Therefore, for comparison with experimental data we use the time-averaged two-dimensional
distribution function, $g_{\tau}(\mathbf{r})$, which represents the probability
of finding a particle at a position ${\mathbf r}$ during a time interval $\tau$,
{\it i.\,e.},
\begin{equation}
  g_{\tau}(\mathbf r)=\frac{1}{N \tau} \left\langle \sum \limits_{i=1}^N \delta(\mathbf r-\mathbf r_i(t)) \right\rangle_{\tau},
  \label{gr}
\end{equation}
where $N$ is the number of particles used in the simulation.

\section{Results and Discussion}
\subsection{Low coverage: $\mathbf{\Theta < 0.1\,\text{ML}}$}
\begin{figure}
  \includegraphics[width=85mm]{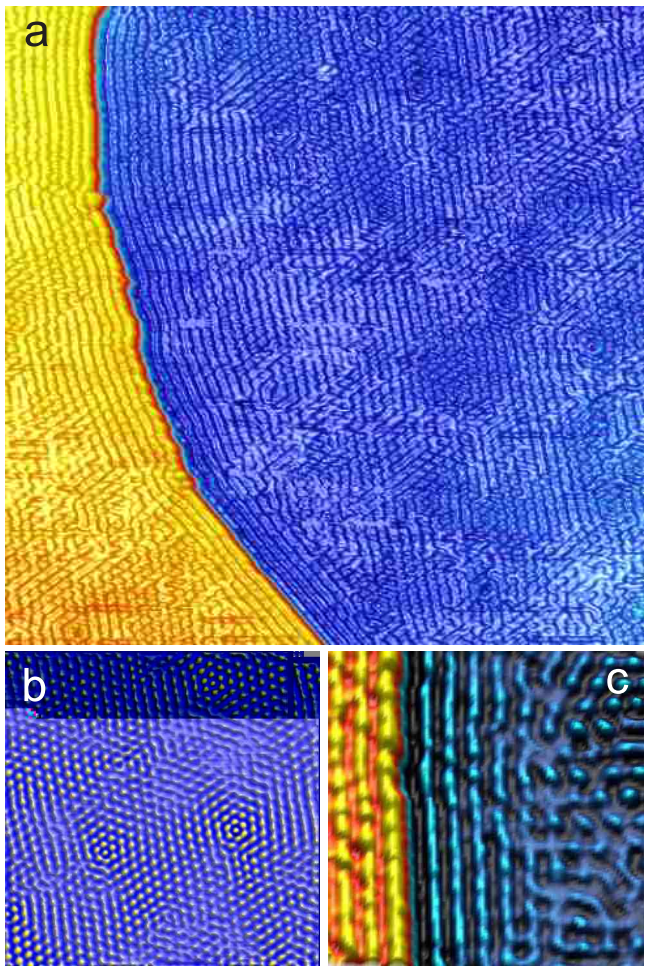}
  \caption[fig1]{(Color online) (a) STM image of two adjacent terraces of
  Ag(111) covered with $0.03$ -- $0.04\,\text{ML}$ Cs deposited at room
  temperature and imaged at $7\,\text{K}$. A monatomically high step separates
  the terraces (voltage $V=200\,\text{mV}$, current $I=0.2\,\text{nA}$, size
  $97\,\text{nm}\times 97\,\text{nm}$). (b) Close-up view of Cs adatoms on
  a terrace ($35\,\text{nm}\times 35\,\text{nm}$). (c) Close-up view of the
  adsorbate arrangement in the vicinity of a step edge
  ($20\,\text{nm}\times 20\,\text{nm}$).}
  \label{fig1}
\end{figure}
Figure \ref{fig1} illustrates the main experimental findings at low coverages.
The STM image in Fig.\,\ref{fig1}(a) shows two adjacent terraces of a Ag(111)
surface covered with $0.03$ -- $0.04\,\text{ML}$ Cs. On terraces we observe
local hexagonal order of Cs atoms. Long-range order is suppressed in regions
where Cs adatoms are no longer resolved as point-like features [Fig.\,\ref{fig1}(b)].
Rather, these regions are characterized by broadened protrusions extending
along the symmetry directions of the adsorbate lattice. From atomically resolved
STM images of the adsorbate lattice a mutual adatom distance of
$(1.5\pm 0.2)\,\text{nm}$ was determined.
At and close to step edges [Fig.\,\ref{fig1}(c)] adatoms tend to form
rows parallel to the step direction separated by $(1.5\pm 0.2)\,\text{nm}$.
Individual Cs adatoms start to be resolved again at distances exceeding
$5\,\text{nm}$ from the step edge.

Below we suggest that the peculiar coexistence of regions with hexagonal order
together with regions in which adatoms are not resolved is likely due to an
adsorbate coverage matching an incommensurate phase. For our kinetic
Monte Carlo simulations several parameters have to be estimated, which are
discussed in the following.
The results of the kinetic Monte Carlo simulations depend on the
transition rates, Eq.\,(\ref{hop}), which are directly coupled
to the parameters of the dipole and the surface state-mediated interactions.
These parameters have to be estimated in advance.
\begin{figure}
  \includegraphics[width=85mm]{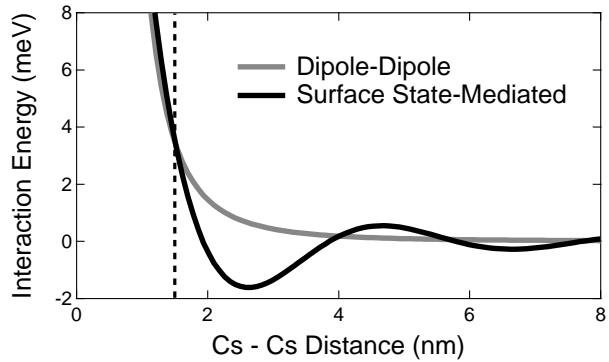}
  \caption[fig2]{Dipole-dipole interaction energy (gray) and surface
  state-mediated interaction energy (black) between Cs adatoms on Ag(111)
  calculated according to Eqs.\,(\ref{dip_int}) and (\ref{SS_Pot}) with
  parameters $p=0.09\,\text{e}\,\text{nm}$, $A_0=0.3$, $\delta_0=\pi/2$,
  $E_0=-0.06\,\text{eV}$, and $k_{\text{F}}=0.813\,\text{nm}^{-1}$. At the
  experimentally observed lattice constant of the adlayer,
  $a\approx 1.5\,\text{nm}$ (dashed line), the surface state-mediated interaction
  and the dipole-dipole interaction energy have similar values.}
  \label{fig2}
\end{figure}

In a first step we estimated the dipole moment of an individual Cs adatom.
Adsorption leads to a considerable charge transfer from an alkali
atom to the substrate. \cite{hpb_87} As a consequence the adsorbed
atom becomes partly ionic and develops a dipole moment which leads to a decrease
of the work function. From the initial linear decrease of the work function
the dipole moment of the adsorbed species may be evaluated. \cite{hpb_87}
The apparent height of the tunneling barrier in STM being related to the work
function, we acquired current-distance curves on clean and Cs-covered Ag(111)
from which we inferred apparent barrier heights of $\approx 5\,\text{eV}$ and
$\approx 4\,\text{eV}$, respectively.
The $1\,\text{eV}$ difference of the apparent barrier heights is in good
agreement with the work function modifications observed on other surfaces covered
with alkali metals. \cite{hpb_89,sal_80,jkr_00} Using a work function change
of $\Delta\Phi\approx 1\,\text{eV}$ we extract a dipole moment $p$ according
to \cite{lds_66}
\begin{equation}
  \epsilon_0\,\Delta\Phi = \text{e}\,p\,n,
\end{equation}
where $-\text{e}$ is the electron charge and $n$ the surface density of alkali
metal atoms. As a result we obtain
$p\approx(0.09\pm 0.03)\,\text{e}\,\text{nm}=(4.3\pm 0.7)\,\text{D}$.

To compare the strength of the dipole-dipole interaction [Eq.\,(\ref{dip_int})]
with the interaction mediated by the surface state [Eq.\,(\ref{SS_Pot})]
we also have to determine the parameters $A_0$, $E_0$, and $\delta_0$. From
previously published results \cite{ewa_98,jre_00,nkn_02,fsi_04} we infer an
amplitude of $A_0\approx 0.3$. The scattering phase shift varies between
$0.3\pi$ and $0.5\pi$ \cite{ewa_98,jre_00,nkn_02,fsi_04} and within this range
our simulations do not depend strongly on the specific choice of $\delta_0$.
We therefore decided to take $\delta_0=\pi/2$. For clean Ag(111) the surface
state binding energy is $E_0\approx -0.06\,{\rm eV}$ below the Fermi energy.
Charge transfer from the Cs layer to the substrate, most likely induces a
shift of the surface state binding energy to higher values. To estimate
this shift, we resort to a similar adsorption system, namely Cu(111)-Cs,
for which a change of the Cu(111) surface state binding energy has been
investigated by photoelectron spectroscopy. \cite{mbr_07} By extrapolating
the photoemission results to low coverages, we find an energy shift of
$\approx 0.01\,\text{eV}$ to higher binding energies for a coverage of
$0.03\,\text{ML}$. However, taking $E_0\approx -0.07\,\text{eV}$
rather than $E_0\approx -0.06\,\text{eV}$ does not alter the results of our
simulations.
For the parameters estimated above we calculated the interaction energies for
the dipole-dipole and the surface state-mediated coupling according to
Eqs.\,(\ref{dip_int}) and (\ref{SS_Pot}). A comparison is shown in
Fig.\,\ref{fig2}. At the adsorbate lattice spacing of $a\approx 1.5\,\text{nm}$
observed in the experiment (dashed line in Fig.\,\ref{fig2}) both interaction
energies are similar. Therefore, in our simulations of the low-coverage adatom
arrangements we take both interactions into account.

To estimate the diffusion barrier height, $E_{\text{D}}$, we observed
single Cs atoms on the cold surface. Upon positioning the tip above the center
of a Cs adatom the feedback loop of the instrument was opened. Next, we
measured the time interval from the moment of freezing the feedback loop until
the current dropped, which signals that the adatom has moved away from its
original position. This procedure resulted in a distribution of time intervals
between $1$ and $15\,\text{s}$. Together with an assumed attempt frequency of
$\nu_0=10^{12}\,\text{s}^{-1}$ (Ref.\,\onlinecite{fic_03}) we estimate a
diffusion barrier height between $16$ and $18\,{\rm meV}$ at $7\,\text{K}$.
We find that the choice of $\nu_0=10^{12}\,\text{s}^{-1}$ and
$E_{\text{D}}\approx 17\,\text{meV}$ matches the time scale of the experiment
quite well. However, a word of caution is necessary in this context. Simulated
time scales vary with the factor $\nu_0\exp(-E_{\text{D}}/\text{k}_{\text{B}}T)$.
Consequently, by modifying $\nu_0$, $E_{\text{D}}$, or both within reasonable
limits, experimentally observed time scales may be reproduced. If experimentally
observed time scales differ from numerically simulated ones by a factor of
more than $10^4$ then it is no longer reasonable to compensate this
discrepancy by an adjustment of $\nu_0$ and $E_{\text{D}}$. \cite{ach_07} In
this case new important physical processes have to be included in the menu of
hopping events of the kinetic Monte Carlo algorithm. \cite{fic_91} It turns
out that the inclusion of the adatom mutual interactions according to
Eqs.\,(\ref{dip_int}) and (\ref{SS_Pot}) already leads to correlated transition
processes.
\begin{figure}
  \includegraphics[width=85mm]{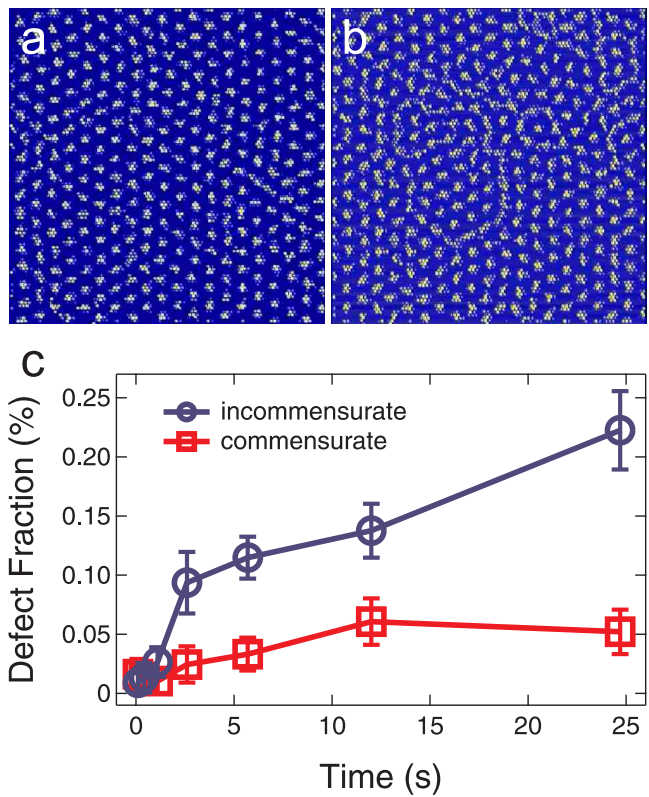}
  \caption[fig3]{(Color online) Density plots of the time-averaged distribution
  function $g_{\tau}(\mathbf{r})$ for $T=7\,\text{K}$, $E_{\text{D}}=17\,\text{meV}$,
  and $\tau=12\,\text{s}$ calculated for particle coverages of (a) $0.037\,\text{ML}$
  and (b) $0.032\,\text{ML}$. The image size corresponds to
  $30\,\text{nm}\times 30\,\text{nm}$. (c) Time evolution of the fraction of
  defects, {\it i.\,e.}, of particles with $5$ and $7$ nearest neighbors for
  an incommensurate (circles) and a commensurate (squares) phase.}
 \label{fig3}
\end{figure}

With these parameters at hand we then addressed the experimentally observed
coexistence of ordered and less ordered regions. To monitor the time evolution
of an initial particle configuration we calculated first the thermal equilibrium
of a particle lattice at $6\,\text{K}$ using canonical Monte Carlo simulations.
In the subsequent kinetic Monte Carlo calculations we set the temperature to
$7\,\text{K}$ as measured in the experiment and analysed the time evolution
of the particle lattice. Figures \ref{fig3}(a) and \ref{fig3}(b) show the
results for particle coverages matching a commensurate [$0.037\,\text{ML}$,
$(\sqrt{27}\times\sqrt{27})R30^{\circ}$] and an incommensurate ($0.032\,\text{ML}$)
superlattice, respectively. The figures show plots of $g_{\tau}(\mathbf{r})$
with $\tau=12\,\text{s}$. Surprisingly, such a slight variation in the coverage
leads to considerably different results.
Whilst in the commensurate adsorption phase [Fig.\,\ref{fig3}(a)]
$g_{\tau}(\mathbf{r})$ is characterized by localized and sharp maxima at sites
of a hexagonal lattice with long-range order, in the incommensurate phase
[Fig.\,\ref{fig3}(b)] $g_{\tau}(\mathbf{r})$ exhibits regions with
maxima which are broader than in the commensurate phase. In particular,
at some positions adjacent maxima overlap and appear as elongated and continuous
lines following close-packed directions of the hexagonal particle lattice.
As a first result we may summarize that particle arrangements that are
incommensurate exhibit enhanced particle mobility.

We performed a Vorono\"{\i} analysis \cite{gvo_07,plu_05} to investigate the
relation between $g_{\tau}(\mathbf{r})$ and particle mobility in more detail.
As a result we obtain the fraction of particles with a number of nearest
neighbors deviating from six. These particles are referred to as defects in
the following.
This analysis thus provides access to local distortions of the hexagonal
symmetry. We define the fraction of these particles according to
$q_{\tau}=N_{\tau}/(\tau N)$ with $N_{\tau}$ denoting the number of defects
recorded during the time interval $\tau$.
Figure \ref{fig3}(c) shows the time evolution of $q_{\tau}$ for the
commensurate (squares) and the incommensurate (circles) phase shown in
Figs.\,\ref{fig3}(a) and \ref{fig3}(b). The fraction increase for
$0<\tau<6\,\text{s}$ is related to thermal relaxation of the initial
particle arrangement at low temperature. The relaxation period is finished
at $\tau\approx 12\,\text{s}$ and $q_{\tau}$ then stays essentially constant
for the commensurate phase. For the incommensurate phase, however, an increase
of $q_{\tau}$ is still observed until $\tau=25\,\text{s}$. Due to costly
computing time we have not performed kinetic Monte Carlo simulations for
larger $\tau$ but we expect $q_{\tau}$ to become constant also for the
incommensurate phase reflecting then thermal equilibrium of the superstructure.
From these simulations we may conclude that once the fraction $q_{\tau}$ does
not exceed $5$ -- $10\,\%$ the superlattice remains stable [Fig.\,\ref{fig3}(a)]
while larger values of $q_{\tau}$ indicate the onset of lattice melting
[Fig.\,\ref{fig3}(b)] or thermal equilibrium which has not been reached.
\begin{figure}
  \includegraphics[width=85mm]{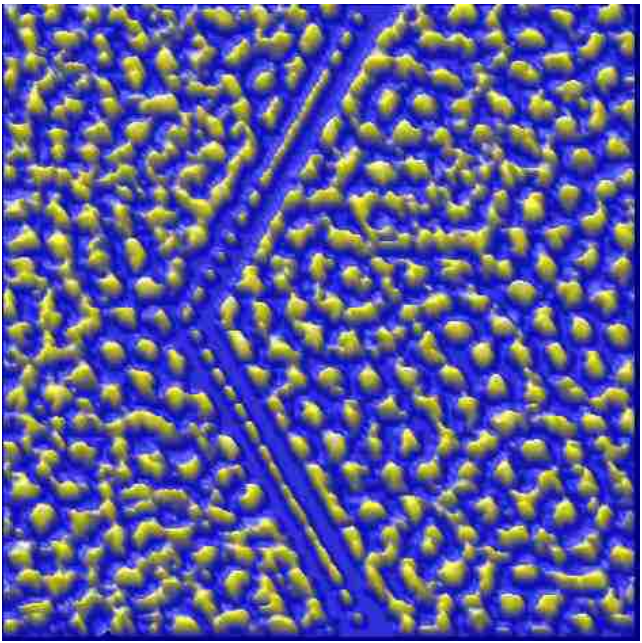}
  \caption[fig4]{(Color online) Density plot of $g_{\tau}(\mathbf{r})$ in the
  vicinity of a $(111)$ (upper part) and a $(100)$ (lower part) step edge. The
  simulation area corresponds to an image size of $30\,\text{nm}\times 30\,\text{nm}$.
  The density plot was generated for $T=7\,\text{K}$, $\tau=12\,\text{s}$,
  and $E_{\text{D}}=17\,\text{meV}$.}
 \label{fig4}
\end{figure}

To model the adsorbate arrangement at step edges, the steps are taken into
account as scatterers of the surface state. We considered a step edge as
an infinitely long and linear chain of Ag atoms each of which gives rise to
a surface state-mediated interaction with Cs adatoms according to Eq.\,(\ref{SS_Pot}).
Integrating the contribution of each scatterer leads to the total interaction
\cite{phy_00,phy_05}
\begin{equation}
  E(r_{\perp}) = B_0\frac{\sqrt{5}E_0}{\pi^2 k_{\text{F}}a_0}\frac{\sin(2k_{\text{F}}r_{\perp} + 2\delta_0 + \pi/4)}{\sqrt{k_{\text{F}}r_{\perp}}^3},
  \label{step}
\end{equation}
where $r_{\perp}$ denotes the distance to the step and $a_0=0.287\,\text{nm}$
the Ag(111) lattice constant. The parameters $E_0$, $k_{\text{F}}$, and
$\delta_0$ are the same that we used for modeling the adlayer structure on
terraces, while the choice of $B_0=4\,\text{meV}$ is in good agreement with
the results obtained for Cu atoms adsorbed on Cu(111). \cite{jre_02}
Figure \ref{fig4} shows the density plot of $g_{\tau}(\mathbf{r})$ for
$\tau=12\,\text{s}$. Two kinds of step edges were included in the simulations,
namely a step edge of the $(111)$ type (upper part of Fig.\,\ref{fig4})
and a $(100)$ step edge (lower part of Fig.\,\ref{fig4}). Deviating from
our simulations of adsorbate structures on terraces we found that the surface
state-mediated interaction according to Eq.\,(\ref{step}) introduced an
additional stabilization of the superlattice. The distribution function,
$g_{\tau}(\mathbf{r})$, exhibits overlapping maxima along the step edge directions
which is in agreement with Cs adatoms showing elongated rows close to step
edges in STM images [Fig.\,\ref{fig1}(c)]. Additionally, maxima of
$g_{\tau}(\mathbf{r})$ become more localized again with larger distances from
the step edge and reflect then the characteristic particle arrangement
on terraces.

According to the Monte Carlo simulations we interpret the experimental
observations as a consequence of an enhanced Cs adatom mobility at the
low-coverage regime discussed in this section. We can identify the origin of
the observed mobility as a property of coverages that match incommensurate
adsorbate phases.

\subsection{Higher coverages: $\mathbf{\Theta} > 0.1\,\text{ML}$}
\begin{figure}
  \includegraphics[width=85mm]{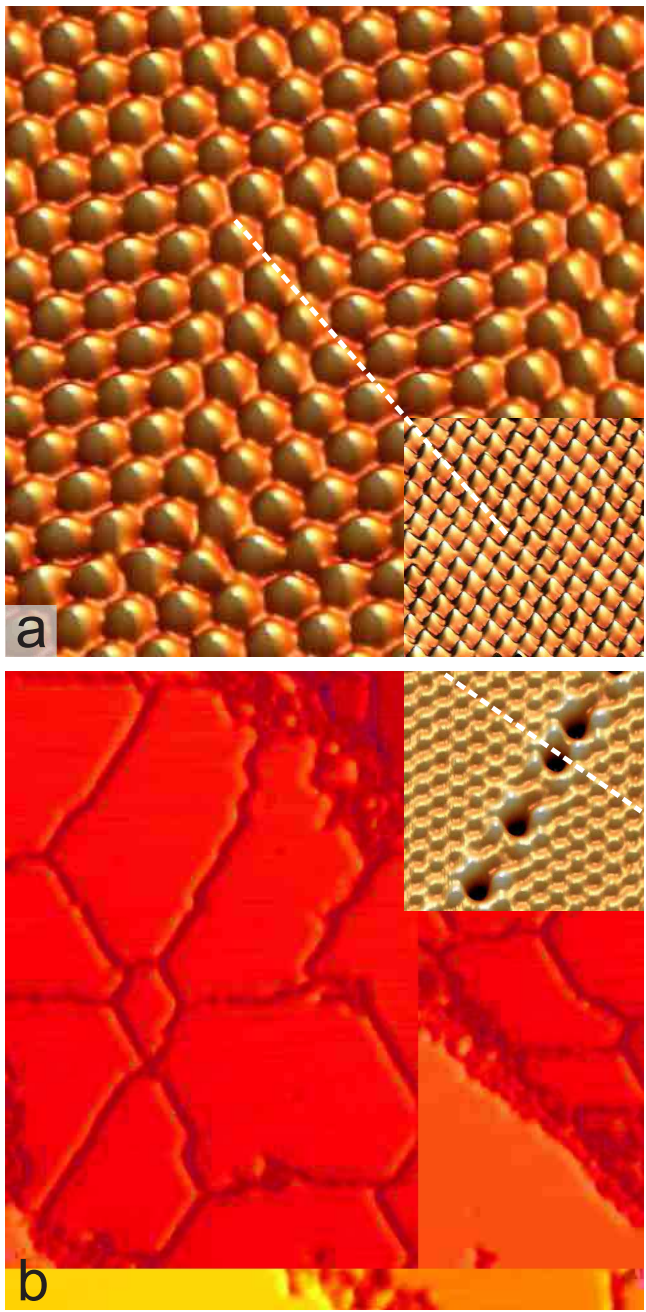}
  \caption[fig5]{(Color online) (a) Quasi-three-dimensional representation of
  constant-current STM image of Ag(111) covered with $0.11\,\text{ML}$ of Cs
  ($V=0.25\,\text{V}$, $I=0.1\,\text{nA}$, $14\,\text{nm}\times 14\,\text{nm}$).
  Cesium adatoms which appear as almost circular protrusions exhibit a mutual
  distance of $0.74\,\text{nm}$. Inset: Atomically resolved Ag(111) lattice.
  The dashed line indicates that adlayer and substrate lattice have the same
  orientation. (b) STM image of Cs-covered Ag(111) at $0.15\,\text{ML}$
  ($V=1.2\,\text{V}$, $I=0.2\,\text{nA}$, $80.4\,\text{nm}\times 80.4\,\text{nm}$).
  Lines on terraces which contain irregularly shaped structures appearing as
  depressions at the applied tunneling voltage are boundaries between translational
  domains of the adsorbate lattice. Inset: Atomically resolved Cs domains.
  While the orientation of the adjacent domains is identical, the lattices
  are translated by half a superlattice constant (see dashed line).}
  \label{fig5}
\end{figure}
A way to stabilize the adsorption lattice, {\it i.\,e.}, to obtain a hexagonal
superstructure with long-range order, is the increase of the coverage. Many
structural analyses of alkali metal films adsorbed on metal surfaces have
previously reported the transition from a disordered phase at very low
coverage to well-ordered adsorbate structures at higher coverages
(Ref.\,\onlinecite{rdd_96} and references therein). An increased coverage leads
to a smaller adatom-adatom distance and will therefore increase the dipole-dipole
repulsion between the adatoms. As a consequence, the probability for a Cs atom
to hop from one adsorption site to an adjacent one is lowered. Our experiment
and calculations corroborate this picture.

Figure \ref{fig5}(a) shows an STM image of Ag(111) covered with $0.11\,\text{ML}$
of Cs. We observe a hexagonal adlayer with long-range order and a
mutual adatom distance of $(0.74\pm 0.04)\,\text{nm}$. This adsorbate
superstructure corresponds to a $(3\times 3)$ commensurate phase and has been
reported before by low-energy electron diffraction. \cite{gle_96} The inset
of Fig.\,\ref{fig5}(a) shows the atomically resolved Ag(111) lattice to
indicate that the adlayer and substrate lattice exhibit the same
orientation [see dashed line in Fig.\,\ref{fig5}(a)].
Regions indicating an enhanced mobility of adatoms are no longer present at
this coverage. Given that the commensurate adsorbate phase is observed
experimentally at a specific coverage, we performed simulations for an adatom
lattice matching the substrate lattice. The superstructure similar to the
one observed in the experiments was obtained as the ground state from canonical
Monte Carlo calculations. The distribution function $g_{\tau}(\mathbf{r})$ for
this coverage was then calculated for a variety of time intervals. Even for
extended intervals, {\it e.\,g.}, $\tau=35\,\text{s}$, no enhancement of the
diffusion processes was observed indicating that the increased dipole-dipole
coupling stabilizes the hexagonal adsorption lattice.

The same behavior was observed at higher coverages. Figure \ref{fig5}(b)
shows an STM image of Ag(111) covered with $0.15\,\text{ML}$ Cs. Again,
no indication of enhanced diffusion of Cs adatoms was found. The Cs adatoms
are arranged in an incommensurate hexagonal superlattice with a mutual distance
of $\approx 0.66\,\text{nm}$ and rotated with respect to the Ag(111) lattice
by $\approx 19^{\circ}$. Moreover, at this coverage domains of hexagonally
ordered Cs atoms were observed. Cesium atomic rows of adjacent domains are
offset by half an adlayer lattice constant [see inset of Fig.\,\ref{fig5}(b)].
At the domain boundaries rows of irregulary shaped structures occur whose
apparent height depends on the tunneling voltage. The formation of domains
may be understood in terms of stress release at the domain boundaries. Similar
observations of translational adsorbate domains have been reported for oxygen
adsorption on W(110) \cite{kjo_93} and on Rh(111). \cite{sma_05}
\begin{figure}
  \includegraphics[width=85mm]{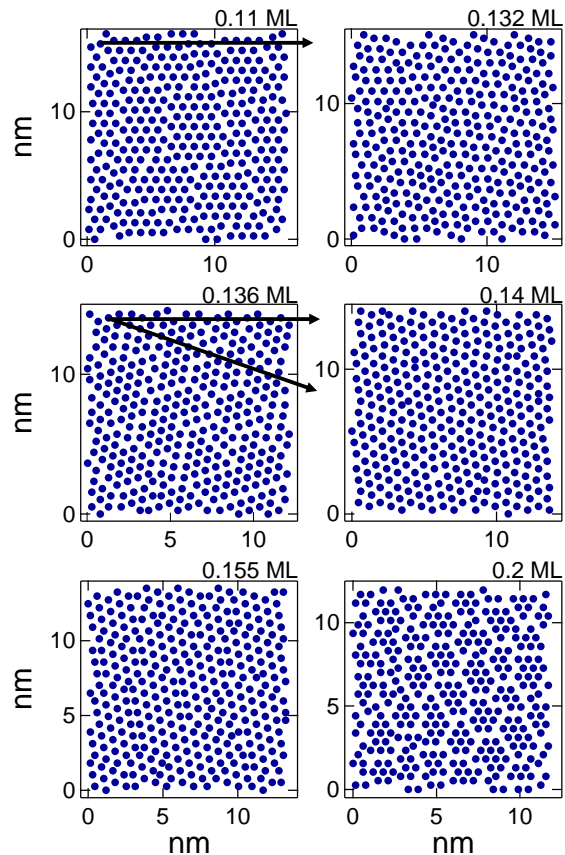}
  \caption[fig6]{Arrangement of $340$ particles according to Kinetic Monte Carlo
  simulations using a time averaging interval of $\tau=12\,\text{s}$ and a
  substrate temperature of $T=7\,\text{K}$. The coverages are indicated at
  the top right of each plot. The horizontal (inclined) arrow indicates the
  crystallographic direction of the substrate (particle lattice). The rotation
  angle is defined as the smaller angle enclosed by the two arrows.}
  \label{fig6}
\end{figure}

\begin{figure}[t]
  \includegraphics[width=85mm]{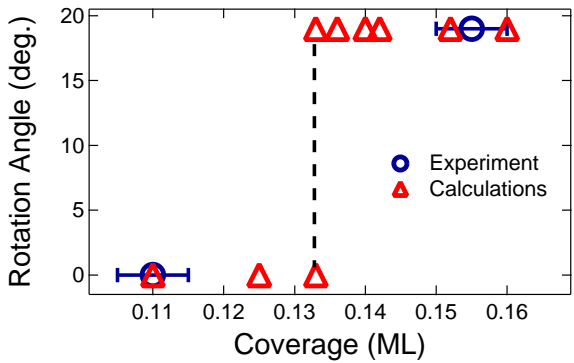}
  \caption[fig7]{(Color online) Rotation angle of the superlattice with respect
  to the substrate lattice. Experimental data are depicted as circles and
  calculated data as triangles. The dashed line indicates a coverage at which
  two phases with different rotation angles coexist.}
  \label{fig7}
\end{figure}

To model these observations within a kinetic Monte Carlo approach we simulated
particle arrangements for coverages ranging between $0.11\,\text{ML}$ and
$0.2\,\text{ML}$ (Fig.\,\ref{fig6}). After an interval of $\tau=12\,\text{s}$
is elapsed the coordinates of $340$ particles with a coverage of $0.11\,\text{ML}$
are given by the upper left plot in Fig.\,\ref{fig6}. This coverage
corresponds to the $(3\times 3)$ commensurate adsorption phase and we see that
the superlattice aligns with the crystallographic direction of the substrate
lattice (depicted as the horizontal arrow). With increasing coverage we observe
that this alignment weakens. At $0.132\,\text{ML}$ some domains are still
oriented along the substrate crystallographic direction while others enclose
an angle of $\approx 20^{\circ}$ indicating that at $0.132\,\text{ML}$
commensurate and incommensurate phases coexist. This coexistence is in agreement
with previous reports based on low-energy electron diffraction. \cite{gle_96}
At $0.136\,\text{ML}$ all particle domains exhibit a rotation angle of
$\approx 20^{\circ}$. This angle is maintained until at $0.2\,\text{ML}$ the
particle lattice is aligned with the substrate lattice again. In particular,
the incommensurate phase observed experimentally at $0.15\,\text{ML}$ with a
rotation angle of $\approx 19^{\circ}$ is well reproduced by our simulations.
We summarize calculated and measured rotation angles as a function of the
coverage in Fig.\,\ref{fig7}.
We notice that our simulations reproduce the experimentally observed rotation
angles for commensurate adsorbate phases. The continuous rotation of incommensurate
phases reported for Ag(111)-Cs \cite{gle_96} and Cu(111)-Cs, \cite{tvh_06}
however, is not concerned here.

Besides the rotation angle of commensurate superstructures the
simulations indicate the origin for domain formation as observed experimentally
at $0.15\,\text{ML}$ [Fig.\,\ref{fig5}(b)]. Increasing the particle density
in the simulations to $0.14\,\text{ML}$ and thus being close to the
$(\sqrt{7}\times\sqrt{7})R19.1^{\circ}$ commensurate phase leads to a particle
arrangement where the domains have almost disappeared in favor of a nearly
homogeneous particle film with long-range hexagonal order. At a particle coverage
of $0.155\,\text{ML}$ the boundaries of adjacent domains overlap leading to
an increase of the particle interaction at the boundaries. A further increase
of the particle density to $0.2\,\text{ML}$ induces a decrease of the domain
size and leads to a similar particle arrangement as observed for $0.136\,\text{ML}$.
As a result we obtain that particle coverages giving rise to incommensurate
phases exhibit the propensity to form domains, while commensurate phases lead
to homogeneous films.

\section{Summary}
Cesium films adsorbed on Ag(111) with submonolayer coverage have been investigated
by scanning tunneling microscopy. The resulting adsorbate arrangements are
modeled by kinetic Monte Carlo simulations. At very low coverages, long-range
hexagonal order of the superstructure does not exist while at higher coverages
a stable and hexagonally ordered adsorbate lattice is observed. Modeling of
the adsorbate arrangement at very low coverages demonstrates the importance
of both dipole-dipole repulsion and surface state-mediated interaction between
Cs adatoms. Long-range hexagonal order is inhibited if the coverage matches
an incommensurate adsorption phase or a disorder-to-order transition. At higher
coverages, the dipole-dipole repulsion between the Cs adatoms becomes large
enough to stabilize hexagonal superlattices. While commensurate adsorption
phases are characterized by a homogeneous adsorbate films, incommensurate
phases exhibit domain-like patterns. The experimentally observed rotation of
commensurate adsorbate lattices is modeled by our calculations.

\section*{Acknowledgements}
We acknowledge funding by the Deutsche Forschungsgemeinschaft through
SFB-TR 24 and SFB 668.


\begin{thebibliography}{nnnsyy}

  \bibitem{rdd_96} R. D. Diehl and R. McGrath,
  Surf.\ Sci.\ Rep.\ {\bf 23}, 43 (1996).

  \bibitem{hpb_89} H. P. Bonzel, A. M. Bradshaw, and G. Ertl (Eds.),
  {\it Physics and chemistry of alkali metal adsorption} (Elsevier, Amsterdam, 1989).

  \bibitem{rgr_85} R. B. Grant and R. M. Lambert,
  Langmuir {\bf 1}, 29 (1985).

  \bibitem{cca_85} C. T. Campbell,
  J.\ Phys.\ Chem.\ {\bf 89}, 5789 (1985).

  \bibitem{gla_88} G. M. Lamble, R. S. Brookes, D. A. King, and D. Norman,
  \prl {\bf 61}, 1112 (1988).

  \bibitem{gle_96} G. S. Leatherman and R. D. Diehl,
  \prb {\bf 53}, 4939 (1996).

  \bibitem{sal_83} S. \AA. Lindgren, L. Walld\'{e}n, J. Rundgren, P. Westrin,
  and J. Neve,
  \prb {\bf 28}, 6707 (1983).

  \bibitem{wcf_88} W. C. Fan and A. Ignatiev,
  \prb {\bf 37}, 5274 (1988).

  \bibitem{zyl_91} Z. Y. Li, K. M. Hock, and R. E. Palmer,
  \prl {\bf 67}, 1562 (1991).

  \bibitem{tvh_06} Th.\ von Hofe, J. Kr\"{o}ger, and R. Berndt,
  \prb {\bf 73}, 245434 (2006).

  \bibitem{sal_88} S.-\AA.\ Lindgren and L. Walld\'{e}n,
  \prb {\bf 38}, 3060 (1988).

  \bibitem{nfi_93} N. Fischer, S. Schuppler, R. Fischer, Th.\ Fauster, and
  W. Steinmann,
  \prb {\bf 47}, 4705 (1993).

  \bibitem{tfa_95} Th.\ Fauster and W. Steinmann,
  in {\it Electromagnetic Waves: Recent Developments in Research},
  edited by P. Halevi, Photonic Probes of Surfaces Vol.\ 2
  (Elsevier, Amsterdam, 1995).

  \bibitem{aca_97} A. Carlsson, B. Hellsing, S.-\AA.\ Lindgren, and L. Walld\'{e}n,
  \prb {\bf 56} 1593 (1997).

  \bibitem{mba_97} M. Bauer, S. Pawlik, and M. Aeschlimann,
  \prb {\bf 55}, 10040 (1997).

  \bibitem{sog_99} S. Ogawa, H. Nagano, and H. Petek,
  \prl {\bf 82}, 1931 (1999).

  \bibitem{bhe_00} B. Hellsing, J. Carlsson, L. Walld\'{e}n, and S.-\AA. Lindgren,
  \prb {\bf 61}, 2343 (2000).

  \bibitem{tch_00} T.-C. Chiang,
  Surf.\ Sci.\ Rep.\ {\bf 39}, 181 (2000).

  \bibitem{mni_01} M. Milun, P. Pervan, and D. P. Woodruff,
  Rep.\ Prog.\ Phys.\ {\bf 65}, 99 (2001).

  \bibitem{jkl_02} J. Kliewer and R. Berndt,
  \prb {\bf 65}, 035412 (2002).

  \bibitem{aka_05} A. K. Kazansky, A. G. Borisov, and J. P. Gauyacq,
  Surf.\ Sci.\ {\bf 577}, 47 (2005).

  \bibitem{cco_05} C. Corriol, V. M. Silkin, D. S\'{a}nchez-Portal, A. Arnau,
  E. V. Chulkov, P. M. Echenique, T. von Hofe, J. Kliewer, J. Kr\"{o}ger, and
  R. Berndt,
  \prl {\bf 95}, 176802 (2005).

  \bibitem{jkr_05} J. Kr\"{o}ger, L. Limot, H. Jensen, R. Berndt, S. Crampin,
  and E. Pehlke,
  Prog.\ Surf.\ Sci.\ {\bf 80}, 26 (2005).

  \bibitem{jko_58} J. Kouteck\'{y},
  Trans.\ Faraday Soc.\ {\bf 54}, 1038 (1958).

  \bibitem{tgr_67} T. B. Grimley,
  Proc.\ Phys.\ Soc.\ (London) {\bf 90}, 751 (1967).

  \bibitem{dne_69} D. N. Newns,
  Phys.\ Rev.\ {\bf 178}, 1123 (1969).

  \bibitem{tei_73} T. L. Einstein and J. R. Schrieffer,
  \prb {\bf 7}, 3629 (1973).

  \bibitem{tts_73} T. T. Tsong,
  \prl {\bf 31}, 1207 (1973).

  \bibitem{kla_78} K. H. Lau and W. Kohn,
  Surf.\ Sci.\ {\bf 75}, 69 (1978).

  \bibitem{hbr_90} H. Brune, J. Wintterlin, G. Ertl, and R. J. Behm,
  Europhys.\ Lett.\ {\bf 13}, 123 (1990).

  \bibitem{mka_96} M. M. Kamna, S. J. Stranick, and P. S. Weiss,
  Science {\bf 274}, 118 (1996).

  \bibitem{ewa_98} E. Wahlstr\"{o}m, I. Ekvall, H. Olin, and L. Walld\'{e}n,
  Appl.\ Phys.\ A {\bf 66}, S1107 (1998).

  \bibitem{jre_00} J. Repp, F. Moresco, G. Meyer, K.-H. Rieder, P. Hyldgaard,
  and M. Persson,
  \prl {\bf 85}, 2981 (2000).

  \bibitem{nkn_02} N. Knorr, H. Brune, M. Epple, A. Hirstein, M. A. Schneider,
  and K. Kern,
  \prb {\bf 65}, 115420 (2002).

  \bibitem{fsi_04} F. Silly, M. Pivetta, M. Ternes, F. Patthey, J. P. Pelz,
  and W.-D. Schneider,
  \prl {\bf 92}, 016101 (2004).

  \bibitem{neg_06} N. N. Negulyaev, V. S. Stepanyuk, L. Niebergall, W. Hergert,
  H. Fangohr, and P. Bruno,
  \prb {\bf 74}, 035421 (2006).

  \bibitem{vss_06} V. S. Stepanyuk, N. N. Negulyaev, L. Niebergall, R. C. Longo,
  and P. Bruno,
  \prl {\bf 97}, 186403 (2006).

  \bibitem{sae_gt} SAES Getters SpA, Viale Italia 77, 20020 Lainate, Italy
  (http://www.saesgetters.com).

  \bibitem{dco_65} D. R. Cox and H. D. Miller,
  {\it The Theory of Stochastic Processes} (Methuen, London, 1965).

  \bibitem{wyo_66} W. M. Young and E. W. Elcock,
  Proc.\ Phys.\ Soc.\ {\bf 89}, 735 (1966).

  \bibitem{fic_91} K. A. Fichthorn and W. H. Weinberg,
  J.\ Chem.\ Phys. {\bf 95}, 1090 (1991).

  \bibitem{fic_03} K. A. Fichthorn and M. Scheffler,
  \prl {\bf 84}, 5371 (2000).

  \bibitem{phy_00} P. Hyldgaard and M. Persson,
  J.\ Phys.\ Condens.\ Matter {\bf 12}, L13 (2000).

  \bibitem{hpb_87} H. P. Bonzel,
  Surf.\ Sci.\ Rep.\ {\bf 8}, 43 (1987).

  \bibitem{sal_80} S.-\AA. Lindgren and L.\ Walld\'{e}n,
  \prb {\bf 22}, 5967 (1980).

  \bibitem{jkr_00} J. Kr\"{o}ger, D. Bruchmann, S. Lehwald, and H. Ibach,
  Surf.\ Sci.\ {\bf 449}, 227 (2000).

  \bibitem{lds_66} L. D. Schmidt and R. Gomer,
  J.\ Chem.\ Phys.\ {\bf 45}, 1605 (1966).

  \bibitem{mbr_07} M. Breitholz, V. Chis, B. Hellsing, S.-\AA. Lindgren,
  and L. Walld\'{e}n,
  \prb {\bf 75}, 155403 (2007).

  \bibitem{ach_07} A. Chatterjee and D. G. Vlachos,
  J.\ Computer-Aided Mater.\ Des.\ {\bf 14}, 253 (2007).

  \bibitem{gvo_07} G. Vorono\"{\i},
  J.\ Reine Angew.\ Math.\ {\bf 133}, 97 (1907).

  \bibitem{plu_05}
  P. Ludwig, S. Kosse, and M. Bonitz,
  \pre {\bf 71}, 046403 (2005).

  \bibitem{phy_05} P. Hyldgaard and T. L. Einstein,
  J.\ Cryst.\ Growth {\bf 275}, e1637 (2005).

  \bibitem{jre_02} J. Repp,
  PhD Thesis (Berlin, 2002).

  \bibitem{kjo_93} K. E. Johnson, R. J. Wilson, and S. Chiang,
  \prl {\bf 71}, 1055 (1993).

  \bibitem{sma_05} S. Marchini, C. Sachs, and J. Wintterlin,
  Surf.\ Sci.\ {\bf 592}, 58 (2005).

\end{thebibliography}
\end{document}